\documentclass[sigconf]{acmart}
\usepackage{wrapfig}
\usepackage{booktabs} % For formal tables
\usepackage{soul}
\usepackage{paralist}
\usepackage{amsmath}
\usepackage{multirow}
\usepackage{tabularx}
\usepackage{balance}
\usepackage{booktabs}
\usepackage{enumitem}
\usepackage[skip=7pt]{caption}
\usepackage[ruled,vlined,linesnumbered]{algorithm2e}
\usepackage{amssymb,amsfonts}
\usepackage{algorithmic}
\usepackage{textcomp}
\usepackage{xcolor}
\usepackage{bm}

\usepackage{dsfont}

\usepackage{tikz-qtree}
\usepackage{tikz-qtree-compat}
\usepackage{dsfont}
\usepackage{amsmath}
\usepackage{graphicx}
\usepackage{subfigure}

\usepackage{color}
\usetikzlibrary{shapes,decorations,arrows,calc,arrows.meta,fit,positioning}
\tikzset{
	-Latex,auto,node distance =1 cm and 1 cm,semithick,
	state/.style ={ellipse, draw, minimum width = 0.7 cm},
	point/.style = {circle, draw, inner sep=0.04cm,fill,node contents={}},
	bidirected/.style={Latex-Latex,dashed},
	el/.style = {inner sep=2pt, align=left, sloped}
}
\newcommand\independent{\protect\mathpalette{\protect\independenT}{\perp}}
\def\independenT#1#2{\mathrel{\rlap{$#1#2$}\mkern2mu{#1#2}}}

\SetArgSty{textnormal}
\setcopyright{rightsretained}
\usepackage{etoolbox}

\apptocmd{\thebibliography}{\small}{}{}
\def\sharedaffiliation{%\end{tabular}
	\begin{tabular}{c}}
	
\begin{document}

\title{Learning Individual Causal Effects from Networked Observational Data}
\author{Ruocheng Guo}
\affiliation{%
	\institution{Arizona State University}
}
\email{rguo12@asu.edu}
\author{Jundong Li}
\affiliation{%
	\institution{University of Virginia}
}
\email{jundong@virginia.edu}
\author{Huan Liu}
\affiliation{%
	\institution{Arizona State University}
}
\email{huan.liu@asu.edu}


% \ccsdesc[300]{Computing methodologies~Causal reasoning and diagnostics}
% \ccsdesc[300]{Computer systems organization~Neural networks}


\copyrightyear{2020}
\acmYear{2020}
\setcopyright{acmcopyright}
\acmConference[WSDM '20]{The Thirteenth ACM International Conference on Web Search and Data Mining}{February 3--7, 2020}{Houston, TX, USA}
\acmBooktitle{The Thirteenth ACM International Conference on Web Search and Data Mining (WSDM '20), February 3--7, 2020, Houston, TX, USA}
\acmPrice{15.00}
\acmDOI{10.1145/3336191.3371816}
\acmISBN{978-1-4503-6822-3/20/02}

\begin{abstract}
    The convenient access to observational data enables us to learn causal effects without randomized experiments.
    This research direction draws increasing attention in research areas such as economics, healthcare, and education. For example, we can study how a medicine (the treatment) causally affects the health condition (the outcome) of a patient using existing electronic health records. 
    To validate causal effects learned from observational data, we have to control confounding bias -- the influence of variables which causally influence both the treatment and the outcome. 
    Existing work along this line overwhelmingly relies on the unconfoundedness assumption that there do not exist unobserved confounders. 
    However, this assumption is untestable and can even be untenable.
    In fact, an important fact ignored by the majority of previous work is that observational data can come with network information that can be utilized to infer hidden confounders.
    For example, in an observational study of the individual-level treatment effect of a medicine, instead of randomized experiments, the medicine is often assigned to each individual based on a series of factors.
    Some of the factors (e.g., socioeconomic status) can be challenging to measure and therefore become hidden confounders. 
    Fortunately, the socioeconomic status of an individual can be reflected by whom she is connected in social networks.
    With this fact in mind, we aim to exploit the network information to recognize patterns of hidden confounders which would further allow us to learn valid individual causal effects from observational data.
    In this work, we propose a novel causal inference framework, the network deconfounder, which learns representations to unravel patterns of hidden confounders from the network information.
    Empirically, we perform extensive experiments to validate the effectiveness of the network deconfounder on various datasets.
\end{abstract}

\maketitle
\section{Introduction}
\begin{figure}[tbh!]
\centering
\begin{minipage}{0.45\textwidth}
\centering
{\includegraphics[width=1\textwidth]{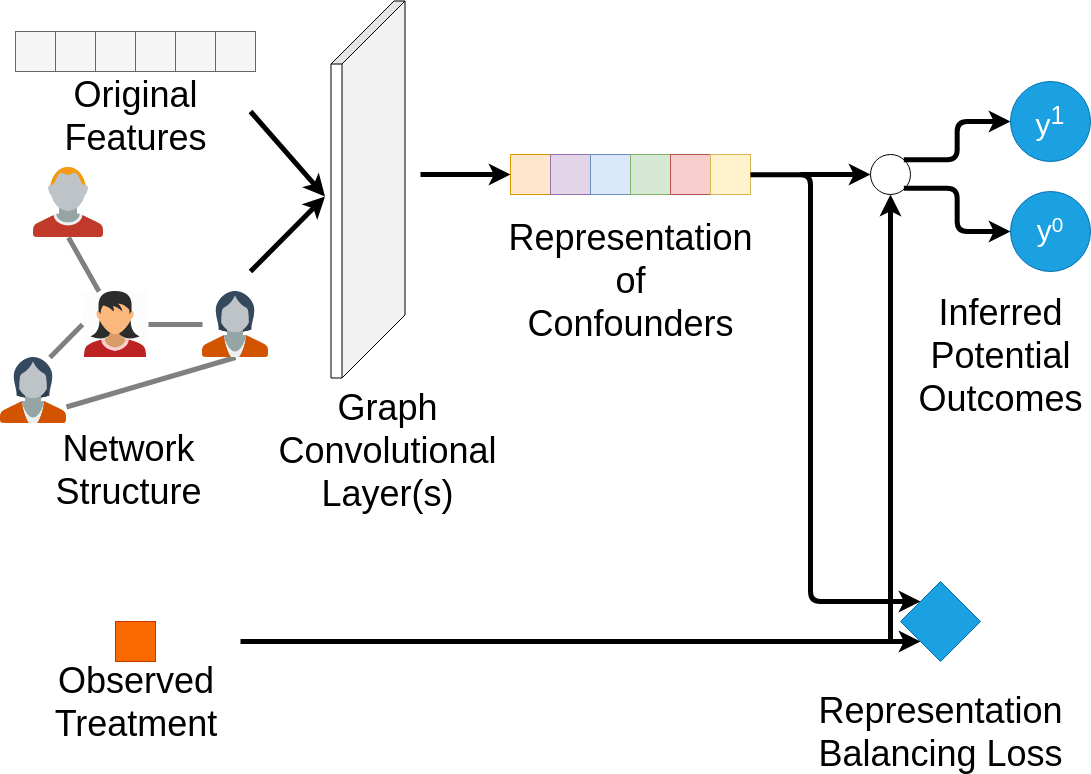}}
\end{minipage}
\centering
\caption{The work flow of the proposed framework network deconfounder}
% \vspace{-0.2in}
\label{fig:intro}
\end{figure}
Recent years have witnessed the rocketing availability of observational data in a variety of highly influential research areas such as economics, healthcare, and education.
Observational data enables researchers to investigate the fundamental problem of learning individual-level causal effects of an interesting treatment (e.g., a medicine) on an important outcome (e.g., the health condition of an individual) without performing randomized controlled trials (RCTs) which can be rather expensive, time consuming, and even unethical~\cite{kallus2018confounding,guo2018survey,cheng2019practical}.
For example, the easy access of a sea of electronic health records ease the studies of the treatment effect of a medicine on patients' health conditions.

Compared to the data collected through RCTs, an observational dataset is often effortless to obtain and comes with a large number of instances and an affluent set features. Meanwhile, the instances are often inherently connected through auxiliary network structures such as a social network connecting users.
Learning individual treatment effects from observational data requires us to handle confounding bias.
We say there exists confounding bias when the causal effect of the treatment on the outcome is distorted by the existence of confounders -- the variables causally influence both the treatment and the outcome.
For example, the poor socioeconomic status of an individual can limit her access to an expensive medicine and have negative impact on her health condition at the same time.
Thus, without controlling the influence of the socioeconomic status, we may overestimate the treatment effect of the expensive medicine.
Controlling the confounding bias is known as the main challenge of learning individual treatment effects from observational data~\cite{rubin1978bayesian,pearl2009causal,guo2018survey}.
%
%\JL{confounders are the challenges? maybe unaware of the confounders are the challenge?}
%
%As a running example, in the observational study of estimating individual treatment effects of a medicine on patients' health conditions, the socioeconomic status of a patient is a confounder as it affects how accessible the medicine is as well as her health condition.
%
To deal with these confounders, a vast majority of existing methods for estimating individual causal effects rely on the strong ignoralbility assumption~\cite{wager2018estimation,schwab2018perfect,hill2011bayesian,johansson2016learning,shalit2017estimating,yao2018representation} that all the confounders are measurable and are embedded in the set of observed features.
%
%% With strong ignorability, it is assumed that the set of features contains all the confounders.
%
As such, these methods often only exploit the observed features to mitigate the confounding bias.
%\JL{do you need to briefly talk about the confounding bias here or you can briefly talk about it before}.
%
In the running example, most of existing efforts try to eliminate the influence of socioeconomic status on the chance to take the medicine and the health condition through controlling the impact of the related proxy variables such as annual income, age, and education.
However, with observational data, the fact is that the causal relations between variables are unknown. As a result, the strong ignoralbility assumption becomes untenable and is likely to be unrealistic due to the existence of hidden confounders~\cite{pearl2009causal}.
Recently, a series of methods have been proposed to leverage the technique of representation learning to relax the strong ignorability assumption.
Nonetheless, they still rely on the assumption that we are able to extract a set of latent features as the set of confounders from observational data using neural networks or factor models~\cite{louizos2017causal,wang2018blessings}.
%

% Traditionally, unobserved confounders are assumed to be those unmeasured features that causally influence the treatments and outcomes of instances at the same time.
%
% Although a wide variety of methods has been proposed to solve the problem of learning individual treatment effects from observational data, 
Despite the existing methods mentioned above, few have recognized the importance of the network structures connecting instances in the task of learning individual treatment effects.
In fact, topology of instances is ubiquitous in various observational data such as a social network of patients, an electrical grid of power stations, and a spatial network of geometric objects, to name a few. 
In addition, when some confounders are notoriously hard to measure, alternatively, we can try to capture their patterns and control their influence by incorporating the underlying network information.
Back to the running example, although the socioeconomic status of an individual is often difficult to be directly measured as a feature, it often can be implicitly represented by social network patterns such as which community she belongs to in a social network~\cite{shakarian2015diffusion}.
Surprisingly, little attention has been paid to utilizing network patterns to mitigate the confounding bias and then improve the precision of estimated individual treatment effects.
To bridge the gap, in this work, we focus on leveraging network patterns as well as the observed features to minimize confounding bias in the task of estimating individual treatment effects.
Note that this work is different from the existing studies on \textit{spillover effect}, also known as \textit{network entanglement} or \textit{interference}~\cite{toulis2018propensity,rakesh2018linked}, where the observed treatment or outcome of an instance may causally influence the outcomes of the connected instances.
In contrast, we focus on the situations where the network structure can be exploited for controlling confounding bias.
For example, a patient's social network patterns can reflect her socioeconomic status but her health condition is not likely to be causally affected by what treatments are assigned to her neighbors in social networks.

To exploit network patterns for mitigating confounding bias, we propose the \textit{network deconfounder}, a novel causal inference framework that captures the influence of hidden confounders from both the original feature space and the auxiliary network information.
Fig.~\ref{fig:intro} illustrates the workflow of the proposed network deconfounder framework.
In particular, the network deconfounder learns representations of confounders by mapping the original features as well as the network structure into a shared latent feature space.
Then the representations of confounders are exploited to control confounding bias and estimate individual treatment effects.
%\JL{could you add one more sentence, what you can do with the learned latent features?}
%

The main contributions of this work are as follows:
\begin{itemize}
    \item We formulate a novel research problem, \emph{learning individual treatment effects from networked observational data}. Note that this problem is different from the studies on interference or spillover effects. We emphasize that digging information about hidden confounders from auxiliary network information can help mitigate confounding bias.
    \item We propose a novel framework for learning individual treatment effects from networked observational data -- \textit{network deconfounder}, which controls confounding bias and estimates individual treatment effects given observational data with auxiliary network information.
    \item We perform extensive experiments to show that the proposed network deconfounder significantly outperforms the state-of-the-art methods for learning individual treatment effects across two semi-synthetic datasets based on real-world social network data.
\end{itemize}

We organize the rest of this paper as follows.
Technical preliminaries and the problem statement of learning individual causal effects from networked observational data is defined in Section 2. 
Section 3 presents the proposed framework, the network deconfounder.
Experiments on semi-synthetic networked observational datasets are presented in Section 4 with discussions.
Section 5 reviews related work.
Finally, Section 6 concludes the paper and visions for the future work.

% \begin{figure}[t]
% 	\centering
% 	\noindent
% 	\begin{minipage}[b]{0.24\textwidth}
% 		\subfigure[Out-sample Performance by varying $\alpha$ from $0.00001$ to $0.01$ .]
% 		{\label{fig:alpha_os}\includegraphics[width=1\textwidth]{./figs/alpha_os.pdf}}
% 	\end{minipage}%\hfill
% 	\begin{minipage}[b]{0.24\textwidth}
% 		\subfigure[Out-sample Performance  as $\lambda$ increases from $0.00001$ to $0.01$.]
% 		{\label{fig:lambda_os}\includegraphics[width=1\textwidth]{./figs/lambda_os.pdf}}
% 	\end{minipage}
% 	\centering
% 	\caption{Parameter studies on the ACIC benchmark with settings: $(\mu_{\kappa1},\mu_{\kappa2})=(3,-1)$, $\mu_{\xi}=2$, $|\mathcal{T}|=10$, $\mu_{\eta}=1.25$.}
% 	\vspace{-0.2in}
% 	\label{fig:param}
% \end{figure}
% from  graph convolution neural network~\cite{defferrard2016convolutional,kipf2016semi}.
%
%
% We summarize existing individual treatment effect estimation approaches  effects either rely on the ignorability assumption which simply ignores the unobserved confounders or learning to approximate them by latent variable focused on eliminating the influence of unobserved confounders through 

\section{Problem Statement}
\label{sec:prob}
\begin{table}[tb]
	\small
	\centering
	\caption{Notations}
    % \vspace{-0.1in}
	\label{tab:notations}
	\begin{tabular}{|c|c|}
		\hline
		\textbf{Symbol} & \textbf{Description} \\\hline
		$\mathbf{x}_i$ & features of the $i$-th instance \\
		$t_i$ & observed treatment of the $i$-th instance\\
		$\mathbf{A}$ & adjacency matrix of the network \\
		$\mathbf{h}_i$ & representation of hidden confounders of instance $i$\\
		%$\bm{t}_i$
		$y_i^F$ or $y_i$ & observed outcome of the $i$-th instance \\
		$y_i^{CF}$ & counterfactual outcome of the $i$-th instance\\
		$y_i^t$ & potential outcome of the $i$-th instance with treatment $t$ \\
		\hline	
% 		$\bm{z}_i$ & post-treatment representation of the $i$-th instance \\ \hline
		$n$ & number of instances \\
		$m$ & dimension of the feature space\\
		$d$ & dimension of the representation space \\
		\hline
	\end{tabular}
\end{table}

In this section, we start with an introduction of the technical preliminaries and then formally present the problem statement of learning individual causal effects from networked observational data.

\noindent\textbf{Notations.} First, we describe the notations used in this work. We denote a scalar, a vector, and a matrix with a lowercase letter (e.g., $t$), a boldface lowercase letter (e.g., $\mathbf{x}$), and a boldface uppercase letter (e.g., $\mathbf{A}$), respectively.
Subscripts signify element indexes (e.g., $\mathbf{x}_i$ and $\mathbf{A}_{i,j}$).
Superscripts of the a potential outcome variable denotes its corresponding treatment (e.g., $y_i^t$). 
Table~\ref{tab:notations} shows a summary of notations that are frequently referred to throughout this work. 

\noindent\textbf{Networked Observational Data.} Then we introduce networked observational data.
In this work, we aim to learn individual treatment effects from networked observational data.
Such data can be represented as $(\{\mathbf{x}_i,t_i,y_i\}_{i=1}^n,\mathbf{A})$ where $\mathbf{x}_i$, $t_i$ and $y_i$ denote the features, the observed treatment, and the observed (factual) outcome of the $i$-th instance, respectively.
The symbol $\mathbf{A}$ signifies the adjacency matrix of the auxiliary network information among different data instances.
Here, we assume that the network is undirected and all the edges share the same weight\footnote{This work can be directly applied to weighted undirected networks. It can also be extended to directed networks using the Graph Convolutional Neural Networks for directed networks~\cite{monti2018motifnet}.}.
Therefore, with the adjacency matrix $\mathbf{A}\in\{0,1\}^{n\times n}$, $\mathbf{A}_{i,j}=\mathbf{A}_{j,i}=1$ ($\mathbf{A}_{i,j}=\mathbf{A}_{j,i}=0$) denotes that there is an (no) edge between the $i$-th instance and the $j$-th instance. %\JL{what about weighted graph?}
We focus on the cases where the treatment variable takes binary values $t\in\{0,1\}$.
Without loss of generality, $t_i=1$ ($t_i=0$) means that the $i$-th instance is under treatment (control).
We also let the outcome variable be a scalar and take values on real numbers as $y\in\mathds{R}$.

Then we introduce the background knowledge of learning individual causal effects.
To define individual treatment effect (ITE), we start with the definition of potential outcomes, which is widely adopted in the causal inference literature~\cite{rubin1978bayesian}\footnote{Note that we only use the concept of potential outcomes, but do not rely on the assumptions that are often adopted along with this concept.}:
\begin{definition}\textbf{Potential Outcomes.}
Given an instance $i$ and the treatment $t$, the potential outcome of $i$ under treatment $t$, denoted by $y^{t}_i$, is defined as the value of $y$ would have taken if the treatment of instance $i$ had been set to $t$.
\end{definition}
Then we are able to provide the formal definition of ITE for the $i$-th instance in the setting of networked observational data as:
\begin{equation}
     \tau_i = \tau(\mathbf{x}_i, \mathbf{A}) = \mathds{E}[y_i^1|\mathbf{x}_i, \mathbf{A}] - \mathds{E}[y_i^0|\mathbf{x}_i, \mathbf{A}]
\end{equation}
Intuitively, ITE is defined as the expected potential outcome of an instance under treatment subtracted by that under control, which reflects how much improvement in the outcome would be caused by the treatment.
Note that with the network information, we are able to go beyond the limited information provided by the features and distinguish two instances with the similar features but different network patterns in the task of learning individual treatment effects.
With ITE defined, we can formulate the average treatment effect (ATE) by taking the average of ITE over the instances as: $ATE = \frac{1}{n}\sum_{i=1}^n \tau_i$.
Finally, we formally present the definition of the problem of learning individual treatment effects from networked observational data as follows:
\begin{definition}\textbf{Learning Individual Treatment Effects from Networked Observational Data.}
Given the networked observational data $(\{\mathbf{x}_i,t_i,y_i\}_{i=1}^n,\mathbf{A})$, we aim to develop a causal inference framework which estimates the ITE of each individual such that an error metric on ITEs is minimized.
\end{definition}

\section{The Proposed Framework}

\subsection{Background}
It is not difficult to find that, in networked observational data, as only one of the two potential outcomes can be observed, the main challenge of learning individual treatment effects is the inference of \textit{counterfactual outcomes} $y_i^{CF} = y_i^{1-t_i}$.
In previous work~\cite{hill2011bayesian,wager2018estimation,johansson2016learning,shalit2017estimating}, with the strong ignorability assumption, controlling observed features is often considered to be enough to eliminate confounding bias.
Formally, strong ignorability can be defined as:
\begin{definition}\textbf{Strong Ignorability.}
With strong ignorability, it is assumed that: (1) the potential outcomes of an instance are independent of whether it receives treatment or control given its features. (2) In addition, for each instance the probability to get treated is larger than 0 and less than 1.
Formally, given the feature space $\mathcal{X}$, the strong ignorability assumption can be presented as:
\begin{equation}
    y^1, y^0 \independent t | \mathbf{x} \;\;\; and  \;\;\; 1 > Pr(t=1|\mathbf{x}) > 0, \forall \mathbf{x} \in \mathcal{X}, t \in \{0,1\}.
\end{equation}
\label{SI}
\end{definition}
It implies that $\mathds{E}[y^t|\mathbf{x}] = \mathds{E}[y|\mathbf{x}, t]$. This is due to the conditional independence between the treatment and the potential outcomes, where $y$ denotes the outcome resulting from the features $\mathbf{x}$ and the treatment $t$.
Intuitively, strong ignorability means we can observe every single feature that describes the difference between the treatment and the control group.
With the strong ignorability assumption, many existing methods~\cite{johansson2016learning,shalit2017estimating,hill2011bayesian,wager2018estimation} boil down the task to learning a machine learning model that approximates the function $f : \mathcal{X} \times \{0,1\} \rightarrow \mathds{R}$ that estimates the expected potential outcomes $\mathds{E}[y|\mathbf{x}, t]$ given features and the treatment.

However, in this work, we consider a more realistic setting where we allow the existence of hidden confounders.
As a result, inferring counterfactual outcomes based on the features and the treatment alone would result in a biased estimator. This can be written as $\mathds{E}[y|\mathbf{x},t]\not = \mathds{E}[y^t|\mathbf{x}]$. This is because the dependencies between the treatment variable and the two potential outcomes are introduced by the hidden confounders.
%

%

% Then we introduce network confounding bias, a new type of confounding bias that has not been defined in the literature.
%
% \begin{definition}\textbf{Network Confounding Bias.}
% We say there exists network confounding bias when the structural pattern of an instance, represented by the adjacency matrix $\mathbf{A}$, influences both its treatment $t$ and outcome $y$. 
% Mathematically, we formulate network confounding bias as:
% \begin{equation}
%     P(y|do(t),\mathbf{x}) \not = 
% \end{equation}

% \end{definition}
\subsection{Network Deconfounder}
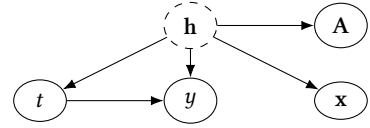
\begin{figure}[tb]
    \centering
    \begin{tikzpicture}
\centering
    % x node set with absolute coordinates
    \node[state] (t) at (-2,0) {${t}$};
    \node[state] (x) at (2,0) {$\mathbf{x}$};

    % y node set relative to x.
    % Locations can be:
    % right,left,above,below,
    % above left,below right, etc
    \node[state] (y) at (0,0) {$y$};
    \node[state,dashed] (h) at (0,1) {$\mathbf{h}$}; 
    \node[state] (a) at (2,1) {$\mathbf{A}$}; 

    % Directed edge
    \path (h) edge (x);
    \path (h) edge (t);
    \path (h) edge (y);
    \path (h) edge (a);
    \path (t) edge (y);
    % Bidirected edge
    % \path[bidirected] (x) edge[bend left=60] (y);
\end{tikzpicture}
    \caption{The causal diagram representing the assumption of the network deconfounder: the network structure represented by adjacency matrix $\mathbf{A}$ along with the observed features $\mathbf{x}$ are proxy variables of the hidden confounders $\mathbf{h}$, which can be utilized to learn representations of hidden confounders. 
    The directed edges signify causal relations, solid circles represent observed variables, and the dashed circle stands for hidden confounders.}
    \label{fig:causal_graph}
\end{figure}

In this subsection, we propose the network deconfounder, a novel framework that addresses the challenges of learning individual treatment effects from networked observational data.
Given the adjacency matrix $\mathbf{A}$, features $\mathbf{x}$, the treatment ${t}$, and the outcome $y$, Fig.~\ref{fig:causal_graph} shows the causal diagram which represents the assumption used in the network deconfounder.
Instead of relying on the strong ignorability assumption, the network deconfounder is based on a weaker assumption that the features and the network structure are two sets of proxy variables of the hidden confounders.
This is a more practical assumption than the strong ignorability assumption in the sense that we do not require the observed features to capture all the information that describes the difference between the treated instances and the controlled ones.
For example, although we cannot directly measure the socioeconomic status of an individual, we can collect features such as age, job type, zip code, and the social network to approximate her socioeconomic status.
Based on this assumption, the proposed network deconfounder attempts to learn representations that approximate hidden confounders and estimate ITE from networked observational data simultaneously.
Unlike eliminating confounding bias based on the observed features alone, leveraging the underlying network structure for controlling confounding bias raises special challenges: (1) instances are inherently interconnected with each other through the network structure and hence their features are not independent identically distributed (i.i.d.) samples from a certain feature distribution, (2) the adjacency matrix of a network is often high-dimensional and can be very sparse ($\mathbf{A}\in\{0,1\}^{n \times n}$).

To tackle these special challenges of controlling confounding bias when network structure information exists, we propose the network deconfounder framework.
The task can be divided into two steps. 
First, we aim to learn representations of hidden confounders by mapping the features and the network structure simultaneously into a shared representation space of confounders.
Then an output function is learned to infer a potential outcome of an instance based on the treatment and the representation of hidden confounders.
Then we present how the two tasks are accomplished by the network deconfounder.

\noindent\textbf{Learning Representation of Confounders.}
In previous work~\cite{johansson2016learning,shalit2017estimating,louizos2017causal}, representation learning techniques have been leveraged for estimating individual level causal effects.
Different from them, the network deconfounder is the first one that is able to utilize auxiliary network information to improve the representation learned toward ITE estimation.
The first component of the network deconfounder is a representation learning function $g$. The function $g$ maps the features and the underlying network into the $d$-dimensional shared latent space of confounders, which can be formulated as $g:\mathcal{X}\times \mathcal{A} \rightarrow \mathds{R}^d$.
We parameterize the $g$ function using Graph Convolutional Networks (GCN)~\cite{defferrard2016convolutional,kipf2016semi}, whose effectiveness have been verified in various machine learning tasks across different types of networked data~\cite{ding2019deep}.
To the best of our knowledge, this is the first work introducing GCN to the task of learning causal effects.
In particular, the representation of confounders of the $i$-th instance is learned through GCN layers. 
Here, for the simplicity of notation, we describe the function $g$ with a single GCN layer.
%
% Note that it can be extended to multiple layers when it is necessary.
%
The representation learning function $g$ is parameterized as:
% \vspace{-2pt}
\begin{equation}
\begin{split}
    \mathbf{h}_i & = g(\mathbf{x}_i,\mathbf{A}) =  \sigma((\hat{\mathbf{A}}\mathbf{X})_i\mathbf{U}), \\
\end{split}
\end{equation}
where $\hat{\mathbf{A}}$ denotes the normalized adjacency matrix, $(\hat{\mathbf{A}}\mathbf{X})_i$ signifies the $i$-th row of the matrix product $\hat{\mathbf{A}}\mathbf{X}$, $\mathbf{U}\in\mathds{R}^{m\times d}$ represents the weight matrix to be learned, and $\sigma$ stands for the ReLU activation function~\cite{glorot2011deep}.
Specifically, with the following notations, $\tilde{\mathbf{A}} = \mathbf{A} + \mathbf{I}_n$ and $\tilde{\mathbf{D}}_{j,j} = \sum_j \tilde{\mathbf{A}}_{j,j}$, the normalized adjacency matrix $\hat{\mathbf{A}}$ can be calculated using the renormalization trick~\cite{kipf2016semi}:
\begin{equation}
    \hat{\mathbf{A}} = \tilde{\mathbf{D}}^{-\frac{1}{2}}\tilde{\mathbf{A}}\tilde{\mathbf{D}}^{-\frac{1}{2}} % = \mathbf{I}_n + \mathbf{D}^{-\frac{1}{2}}\mathbf{A}\mathbf{D}^{-\frac{1}{2}}
\end{equation}
We can compute $\hat{\mathbf{A}}$ in a pre-processing step to avoid repeating the computation.
Then the weight matrix $\mathbf{U}\in\mathds{R}^{m\times d}$ along with the ReLU activation function maps the input signal into the low-dimensional representation space.
Note that more than one GCN layers can be stacked to catch the non-linear relations between hidden confounders and the input data.

\noindent\textbf{Inferring Potential Outcomes.}
Then we introduce the second component of network deconfounder, namely the output function $f:\mathds{R}^d \times \{0,1\} \rightarrow \mathds{R}$.
The function $f$ maps the representation of hidden confounders as well as a treatment to the corresponding potential outcome. 
%
% The network deconfounder is a representation learning framework which controls network confounders computed through graph convolutional layers~\cite{kipf2016semi,defferrard2016convolutional} as:
With $\mathbf{h}_i\in \mathds{R}^{d}$ denoting the representation of the confounders of the $i$-th instance and $t\in\{0,1\}$ signifying the treatment, to infer the corresponding potential outcome, the output function $f$ is defined as:
\begin{equation}
    f(\mathbf{h}_i,t) = 
    \begin{cases}
    f_1(\mathbf{h}_i)\text{ if }t = 1\\
    f_0(\mathbf{h}_i)\text{ if }t = 0\\
    \end{cases},
\end{equation}
where $f_1$ and $f_0$ are the output functions for treatment $t=1$ and $t=0$.
Specifically, we parameterize the output functions $f_1$ and $f_0$ using $L$ fully connected layers followed by a regression layer as:
\begin{equation}
\begin{split}
    f_1 = \mathbf{w}^1\sigma(\mathbf{W}_L^1...\sigma(\mathbf{W}_1^1\mathbf{h}_i)), \\
    f_0 = \mathbf{w}^0\sigma(\mathbf{W}_L^0...\sigma(\mathbf{W}_1^0\mathbf{h}_i)),
\end{split}
\end{equation}
where $\mathbf{h}_i$ is the representation of the $i$-th instance's confounders (output of the $g$ function), $\{\mathbf{W}_l^t\}, l=1,...,L$ denote the weight matrices of the fully connected layers, and $\mathbf{w}^t$ is the weight for the regression layers.
The bias terms of the fully connected layers and the output regression layer are dropped for simplicity of notation.
We can either set $t=t_i$ to infer the observed factual outcome $y_i^{CF}$ or $t=1-t_i$ to estimate the counterfactual outcome.

%
% Then through a set of neural network layers, we map the representation of the $i$-th instance to one of its potential outcomes as:

With the two components of the network deconfounder formulated, given the features of the $i$-th instance $\mathbf{x}_i$, the treatment $t$, and the adjacency matrix $\mathbf{A}$, we can infer the potential outcome as:
\begin{equation}
    \hat{y}_i^{t} = f(g(\mathbf{x}_i,\mathbf{A}),t),
\end{equation}
where $\hat{y}_i^{t}$ denotes the inferred potential outcome of instance $i$ corresponding to treatment $t$ by the network deconfounder framework.

\noindent\textbf{Objective Function.}
Then, we introduce the three essential components of the loss function for the proposed network confounder.

\textit{Factual Outcome Inference.} First, we aim to minimize the error in the inferred factual outcomes.
This leads to the first component of the loss function, the mean squared error in the inferred factual outcomes:
\begin{equation}
    \frac{1}{n}\sum_{i=1}^N(\hat{y}_i^{t_i}-y_i)^2.
\end{equation}
% where $w_i$ denotes the sample weight of the $i$-th instance. We set the sample weight of the $i$-th instance as its inverse propensity score~\cite{guo2018survey} as $w_i=\frac{t_i}{Pr(t_i=1)}+\frac{1-t_i}{1-Pr(t_i=1)}$.
%
% Following~\cite{shalit2017estimating}, we estimate the propensity score as the proportion of instances receiving the treatment to compensate for the difference in the number of instances receiving the treatment and the control ($Pr(t_i=1)=\frac{|\{i|t_i=1\}|}{n}$).

\textit{Representation Balancing.}
Minimizing the error in the factual outcomes ($y_i$) does not necessarily mean that the error in the counterfactual outcomes ($y_i^{CF}$) is also minimized.
In other words, in the problem of learning ITE from networked observational data, we essentially confront the challenge of distribution shift~\cite{johansson2016learning,shalit2017estimating}.
In particular, the network deconfounder would be trained on the conditional distribution of factual outcomes $Pr(y_i|\mathbf{x}_i,\mathbf{A},t_i)$ but the task is to infer the conditional distribution of counterfactual outcomes $Pr(y_i^{CF}|\mathbf{x}_i,\mathbf{A},1-t_i)$. 
In~\cite[Lemma 1.]{shalit2017estimating}, the authors have shown that the error in the inferred counterfactual outcomes is upperbounded by a weighted sum of (1) the error in the inferred factual outcomes; and (2) an integral probability metric (IPM) measuring the difference between the distributions the treated instances and the controlled instances in terms of their confounder representations.
Therefore, besides the error in inferred factual outcomes, we also aim to minimize the IPM measuring the how different the treatment group and the control group are regarding their distributions of confounders' representations.
With $P(\mathbf{h})=Pr(\mathbf{h}|t_i=1)$ and $Q(\mathbf{h})=Pr(\mathbf{h}|t_i=0)$ being the empirical distributions of representation of hidden confounders, we let $\rho_\mathcal{Z}(P,Q)$ denote the IPM defined in the functional space $\mathcal{Z}$ which measures the divergence between the two distributions of confounders' representations.
Assuming that $\mathcal{Z}$ denotes the functional space of 1-Lipschitz functions, the IPM reduces to the Wasserstein-1 distance which is defined as:
\begin{equation}
\label{eq:w_dist}
    \rho_\mathcal{Z}(P,Q) = \underset{k\in\mathcal{K}}{inf}\int_{\mathbf{h}\in \{\mathbf{h}_i\}_{i:t_i=1}} ||k(\mathbf{h})-\mathbf{h}||P(\mathbf{h})d\mathbf{h}
\end{equation}
where $\mathcal{K} = \{k|k:\mathds{R}^d\rightarrow \mathds{R}^d \; s.t. \; Q(k(\mathbf{h}))=P(\mathbf{h})\}$ denotes the set of push-forward functions that can transform the representation distribution of the treated ($P(\mathbf{h})$) to that of the controlled ($Q(\mathbf{h})$). 
By minimizing $\alpha\rho_\mathcal{Z}(P,Q)$, we approximately minimize the divergence between the distributions of confounders' representations, where $\alpha\ge 0$ signifies the hyperparameter controlling the trade-off between penalizing the imbalance of confounders' representations and the other penalty terms in the loss function of the network deconfounder.
We adopt the efficient approximation algorithm proposed by~\cite{cuturi2014fast} to compute the Wasserstein-1 distance in Eq.~\eqref{eq:w_dist} and its gradients against the model parameters for training the network deconfounder.
%

%
% \begin{equation}
%     \alpha IPM_G(\{\mathbf{h}_i\}_{i:t_i=1},\{\mathbf{h}_j\}_{j:t_j=0}).
% \end{equation}
%

%

%
% Due to the availability of the counterfactual outcomes, we cannot directly minimize the difference $IPM_G(Pr(y_i|\mathbf{x}_i,\mathbf{A},t_i)||Pr(y_i|\mathbf{x}_i,\mathbf{A},1-t_i))$

\textit{$\ell_2$ Regularization.} Third, we let $\bm{\theta}$ signify the vector of the model parameters of the network deconfounder.
Then a squared $\ell_2$ norm regularization term on the model parameters - $\lambda||\bm{\theta}||_2^2$, is added to mitigate the overfitting problem, where $\lambda \ge 0$ denotes the hyperparamter controlling the trade-off between the $\ell_2$ regularization term and the other two terms.

Formally, we present the objective function of the network deconfounder as:
\begin{equation}
\label{eq:obj}
    \mathcal{L}(\{\mathbf{x}_i,t_i,y_i\}_{i=1}^n,\mathbf{A}) = \frac{1}{n}\sum_{i=1}^n (\hat{y}_i^{t_i}-y_i)^2  + \alpha \rho_\mathcal{Z}(P,Q) + \lambda ||\bm{\theta}||_2^2,
\end{equation}
% In addition, $||\bm{\theta}||_2^2$ signifies the squared $\ell_2$ regularization term. Finally, we penalize the imbalance between the two treatment groups in the representation space by minimizing $IPM_G(\{\mathbf{h}_i\},\{\mathbf{h}(\mathbf{x}_i,\mathbf{A})\})$, which denotes the empirical integral probability metric defined by the functional space $G$~\cite{shalit2017estimating}.
% %
% The non-negative hyperparameters, $\lambda$ and $\alpha$ control the trade-off between the prediction accuracy on the observed outcomes, the strength of $L_2$ regularization and the penalty on imbalance in the representation space. 
\section{Experiments}

\subsection{Dataset Description}

It is notoriously hard to obtain ground truth of ITEs because in most if not all cases, we can only observe one of the potential outcomes.
For example, a patient can only choose to take the medicine or not to take it, but not both. So we can only observe the outcome resulting from her choice.
However, we need benchmark datasets that provide ground truth of ITEs such that we can compare different methods that estimate ITEs with networked observational data.
To resolve this problem, we follow the existing literature~\cite{johansson2016learning,shalit2017estimating,louizos2017causal,schwab2019learning} to create semi-synthetic datasets.
In particular, we introduce two benchmark datasets for the task of learning ITEs from networked observational.
These datasets are semi-synthetic in the sense that they are based on features and network structures collected from real-world sources. Then we synthesize treatments, and outcomes for the task of learning ITEs from networked observational data in the presence of hidden confounders.

\noindent\textbf{BlogCatalog.}
BlogCatalog\footnote{https://www.blogcatalog.com/} is an online community where users post blogs.
In the dataset, each instance is a blogger. Each edge signifies the social relationship (friendship) between two bloggers.
The features are bag-of-words representations of keywords in bloggers' descriptions.
We extend the BlogCatalog dataset used in~\cite{li2015unsupervised,li2019adaptive} by synthesizing (a) the outcomes -- the opinions of readers on each blogger; and (b) the treatments -- whether contents created by a blogger receive more views on mobile devices or desktops. 
Similar to the News dataset used in the previous work~\cite{johansson2016learning,schwab2018perfect,schwab2019learning}, we make the following assumptions: (1) Readers either read on mobile devices or desktops. We say a blogger get treated (controlled) if her blogs are read more on mobile devices (desktops). (2) Readers prefer to read some topics from mobile devices, others from desktops.
(3) A blogger and her neighbors' topics causally influence her treatment assignment.
(4) A blogger and her neighbors' topics also causally affect readers' opinions on them.
Here, we aim to study the individual treatment effect of receiving more views on mobile devices (than desktops) on readers' opinions.
To synthesize treatments and outcomes in accordance to the assumptions mentioned above, we first train a LDA topic model~\cite{blei2003latent} on a large set of documents.
Then, two centroids in the topic space are defined as follows: (i) we randomly sample a blogger and let the topic distribution of her description be the centroid of the treated instances, denoted by $r_1^c$. (ii) The centroid of the controlled, $r_0^c$, is set to be the mean of the topic distributions of all the bloggers' descriptions.
Then we introduce how the treatments and outcomes are synthesized based on the similarity between the topic distribution of a blogger's description and the two centroids.
With $r(\mathbf{x}_i)$ denoting the topic distribution of the $i$-th blogger's description, we model the device preference of the readers of the $i$-th blogger's content as:
\begin{equation}
\begin{split}
    Pr(t&=1|\mathbf{x}_i,\mathbf{A}) = \frac{\exp(p_1^i)}{\exp(p^i_1)+\exp(p_0^i)}; \\ 
p_1^i & = \kappa_1 r(\mathbf{x}_i)^T r_1^c+\kappa_2\sum_{j\in\mathcal{N}(i)}r(\mathbf{x}_j)^Tr_1^c \\ &=  \kappa_1 r(\mathbf{x}_i)^T r_1^c+\kappa_2(\mathbf{A}r(\mathbf{x}_j))^Tr_1^c;\\
p_0^i & = \kappa_1 r(\mathbf{x}_i)^T r_0^c+\kappa_2\sum_{j\in\mathcal{N}(i)}r(\mathbf{x}_j)^Tr_0^c \\ & =  \kappa_1 r(\mathbf{x}_i)^T r_0^c+\kappa_2(\mathbf{A}r(\mathbf{x}_j))^Tr_0^c,\\
\end{split}
\end{equation}
where $\kappa_1, \kappa_2 \ge 0$ signifies the magnitude of the confounding bias resulting from a blogger's topics and her neighbors' topics, respectively.
When $\kappa_1=0, \kappa_2=0$ the treatment assignment is random and the greater the values $\kappa_1$ and $\kappa_2$ are, the more significant the influence of a blogger's topics and her neighbors' topics on the device preference is.
%
%
% It worth to mention that the network structure implicitly influences the features of each instance through social influence and Homophily, although it is not an explicit input of the propensity score $Pr(t=1|\mathbf{x}_i)$, 
%
Then the factual outcome and the counterfactual outcome of the $i$-th blogger are simulated as:
\begin{align}
    y^F(\mathbf{x}_i)= y_i=C(p_0^i+t_i p_1^i)+\epsilon; \\
    y^{CF}(\mathbf{x}_i)=C[p_0^i+(1-t_i) p_1^i]+\epsilon,
\end{align}
where $C$ is a scaling factor and the noise is sampled as $\epsilon\sim\mathcal{N}(0,1)$.
In this work, we set $C = 5, \kappa_1=10, \kappa_2 \in \{0.5,1,2\}$.
Note that the outcomes of an individual are not influenced by the treatment assignment or outcomes of their neighbors, therefore, there is no interference or spillover effect in this scenario.

\begin{figure}
\centering
\includegraphics[width=0.48\textwidth]{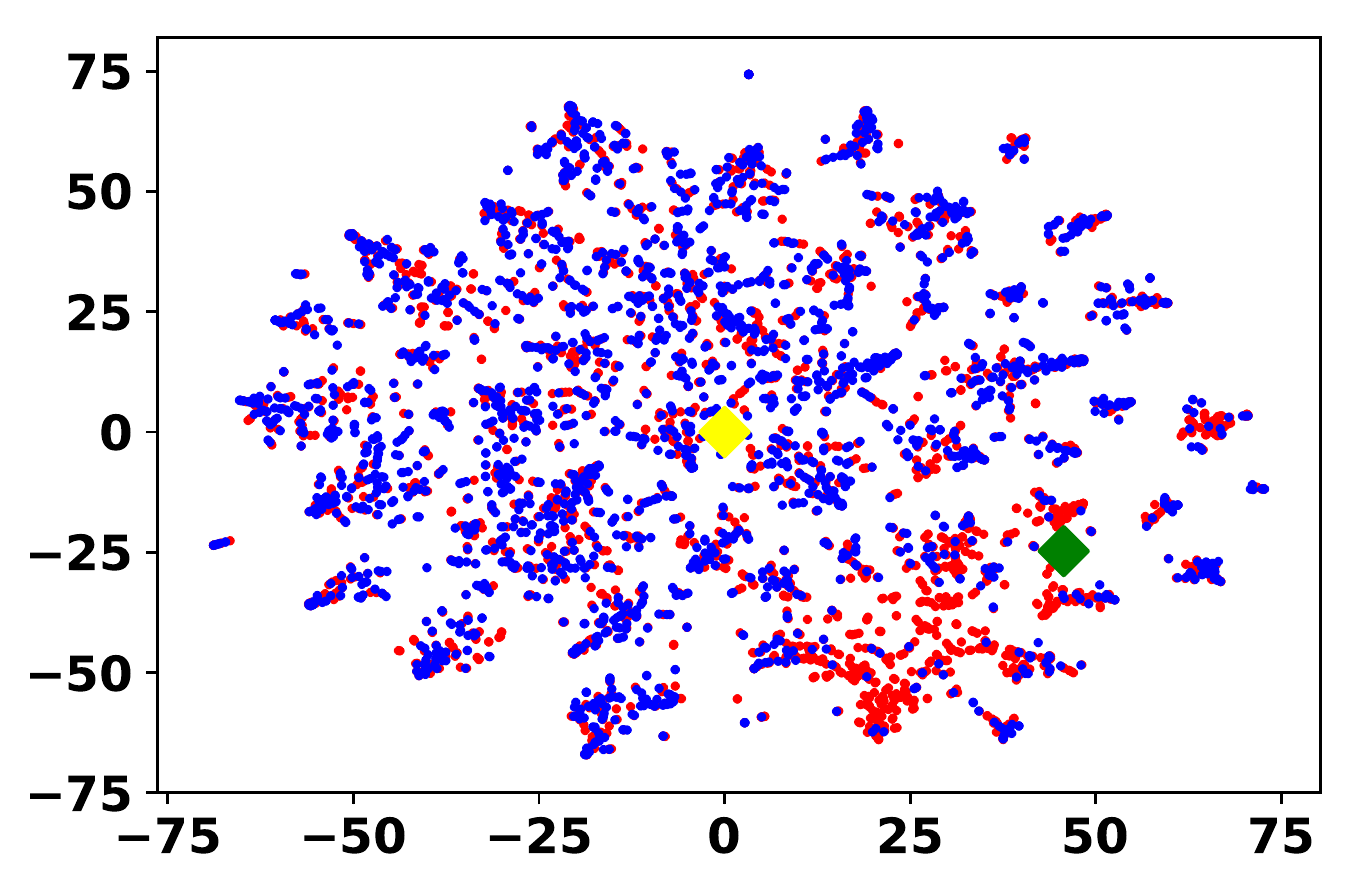}
\caption{Distribution of the treated (red) and control (blue) instances in the LDA topic space. The green and yellow diamonds signify the two centroids $r_1^c$ and $r_0^c$ for the two treatment groups.} \label{fig:bctsne}
\end{figure}

In the experiments, 50 LDA topics are learned from the training corpus. 
Then we reduce the vocabulary by taking the union of the most frequent 100 words from each topic. By doing this, we end up with 2,173 bag-of-words features.
We perform the aforementioned simulation 10 times for each setting of $\kappa_2$.
Figure~\ref{fig:bctsne} shows the distribution of topics in one of the simulations, which is projected to two-dimensional space through the visualization technique TSNE~\cite{vanDerMaaten2008}.
We observe that there are more treated instances (red dots) near the centroid $r_1^c$ (green diamond) and more control instances (blue dots) close to the centroid $r_0^c$ (yellow diamond). In addition, a significant shift from the centroids can be perceived which shows the impact of the network structure.

\noindent\textbf{Flickr.} Flickr\footnote{https://www.flickr.com} is an online social network where users share images and videos. In this dataset, each instance is a user and each edge represents the social relationship (friendship) between two users.
The features of each user represent a list of tags of interest.
We adopt the same settings and assumptions as we do for the BlogCatalog dataset.
Thus, we also study the individual-level causal effects of being viewed on mobile devices on readers' opinions on the user.
In particular, we also learn 50 topics from the training corpus using LDA and concatenate the top 25 words of each topic.
Thus, we reduce the data dimension to 1,210.
We maintain the same settings of parameters as the BlogCatalog dataset ($C=5, \kappa_1=10$ and $\kappa_2\in\{0.5,1,2\}$).

\begin{table}
\caption{Dataset Description: BC stands for BlogCatalog}\label{tab:datasets}
\centering
\small
\begin{tabular}{|c|c|c|c|c|c|c|c|}
\hline
 & Instances & Edges & Features & $\kappa_2$ & ATE Mean & STD\\
\hline
\multirow{3}{*}{BC} & \multirow{3}{*}{5,196} & \multirow{3}{*}{173,468} & \multirow{3}{*}{8,189} & 0.5 & 4.366 & 0.553\\
&&&& 1 & 7.446 & 0.759 \\
 &&&& 2 & 13.534 & 2.309  \\ \hline
\multirow{3}{*}{Flickr} & \multirow{3}{*}{7,575} & \multirow{3}{*}{239,738} & \multirow{3}{*}{12,047} & 0.5 & 6.672 & 3.068\\
&&&& 1 & 8.487 & 3.372 \\
&&&& 2 & 20.546 & 5.718 \\ \hline
\end{tabular}
\end{table}

In Table~\ref{tab:datasets}, we present a summary of the statistics of the semi-synthetic datasets described in this subsection.
The average and standard deviation of the ATEs are calculated over the 10 runs under each setting of parameters.

\subsection{Experimental Settings}

Following the original implementation of GCN~\cite{kipf2016semi}\footnote{https://github.com/tkipf/gcn}, we train the model with all the training instances along with the complete adjacency matrix of the auxiliary network information.
ADAM~\cite{kingma2014adam} is the optimizer we use to minimize the objective function of the network deconfounder (Eq.~\eqref{eq:obj}).
We randomly sample $60\%$ and $20\%$ of the instances as the training set and validation set and let the remaining be the test set. We perform 10 times of random sampling for each simulation of the datasets and report the average results.
Grid search is applied to find the optimal combination of hyperparameters for the network deconfounder.
In particular, we search learning rate in $\{10^{-1},10^{-2},10^{-3},10^{-4}\}$, the number of output layers in $\{1,2,3\}$, dimensionality of the outputs of the GCN layers and the number of hidden units of the fully connected layers in $\{50,100,200\}$, $\alpha$ and $\lambda$ in $\{10^{-3},10^{-4},10^{-5},10^{-6}\}$.
For the baselines, we adopt their default settings of hyperparameters.

Note that, in this work, we consider the scenarios where each individual's potential outcomes are not influenced by the observed treatments or outcomes of others in the network, i.e., there is no interference or spillover effect.
At the same time, the auxiliary network is utilized as a source of information to help us learn better representations of confounders.
Also note that the proposed network deconfounder framework is the first framework which incorporates auxiliary network information to learn better representations for controlling confounding bias and estimating individual treatment effects.
Therefore, there does not exist baseline methods that can naturally incorporate the auxiliary network information.
But we can concatenate the corresponding row of adjacency matrix to the original features to enable the baselines to utilize the network information.
However, due to the issues of high dimensionality and sparsity, we find such an approach cannot improve baselines' performance.
Then we describe the baseline methods which represent the state-of-the-art methods for the task of learning ITEs from observational data.

    \noindent\textbf{Counterfactual Regression (CFR)~\cite{shalit2017estimating}.} CFR is based on the strong ignorability assumption.
    It learns representations of confounders by mapping the original features into a latent space.
    CFR is trained by minimizing the error in inferred factual outcomes and tries to minimize the imbalance of confounders' representations between the treated and the controlled.
    Following~\cite{shalit2017estimating}, two types of representation balancing penalties are considered: the Wasserstein-1 distance (CFR-Wass) and the maximum mean discrepancy (CFR-MMD).

    \noindent\textbf{Treatment-agnostic Representation Networks (TARNet)~\cite{shalit2017estimating}.} TARnet is a variant of CFR which comes without the representation balancing penalty term.
    
    \noindent\textbf{Causal Effect Variational Autoencoder (CEVAE)~\cite{louizos2017causal}.} CEVAE is a deep latent-variable model which estimates ITEs via modeling the joint distribution $P(\mathbf{x},t,y,\mathbf{h})$.
    It learns representations of confounders as Gaussian distributions. Then through variational inference, it is trained by maximizing the variational lower bound of the graphical model representing the causal relations between the four variables: the features, the treatment, the outcome and the confounders.

    \noindent\textbf{Causal Forest~\cite{wager2018estimation}.} Causal Forest is an extension of Breiman's random forest~\cite{breiman2001random} for estimating heterogenous treatment effects in subgroups. Here, we treat the heterogenous treatment effect estimated by causal forest of a subgroup as the ITE of each instance in the subgroup.
    It works with the strong ignorability assumption.

\noindent\textbf{Bayesian Additive Regression Trees (BART)~\cite{hill2011bayesian}.} BART is a Bayesian regression tree based ensemble model which is widely adopted in the literature of causal inference. It is also based on the strong ignorability assumption.

Two widely used evaluation metrics, the Rooted Precision in Estimation of Heterogeneous Effect ($\sqrt{\epsilon_{PEHE}}$) and Mean Absolute Error on ATE ($\epsilon_{ATE}$), are adopted by this work.
Formally, they are defined as:
\begin{equation}
\begin{split}
       \sqrt{\epsilon_{PEHE}} = \sqrt{\frac{1}{n}\sum_{i=1}(\hat{\tau}_i-\tau_i)^2}, \\
        \epsilon_{ATE} = |\frac{1}{n}\sum_{i=1}(\hat{\tau}_i) - \frac{1}{n}\sum_{i=1}(\tau_i)|,
\end{split}
\end{equation}
where $\hat{\tau}_i = \hat{y}_i^1 - \hat{y}_i^0$
% where\begin{equation}
%   \hat{\tau}_i = \begin{cases} y_i^1 - \hat{y}_i^0 \\
%   \hat{y}_i^1 - y_i^0\\
%     \end{cases}
% \end{equation} 
 and $\tau_i = y_i^1 - y_i^0$ denote the inferred ITE and the ground truth ITE for the $i$-th instance.

\subsection{Results}

\noindent\textbf{Effectiveness.} First, we compare the effectiveness of the proposed framework, the network deconfounder, with the aforementioned state-of-the-art methods.
Table~\ref{tab:res1} summarizes the empirical results on the BlogCatalog and Flickr datasets with $C = 5, \kappa_1 = 10$ and $\kappa_2 \in\{0.5,1,2\}$.
We summarize the observations from these experimental results as follows:
\begin{itemize}
    \item The proposed network deconfounder framework consistently outperforms the state-of-the-art baseline methods on the semi-synthetic datasets with treatments and outcomes generated under various settings. We also perform one-tailed T-test to verify the statistical significance. The results indicate that the network deconfounder achieves significantly better estimations on individual treatment effects with a significant level of 0.05.
    \item With the capability to recognize the patterns of hidden confounders from the network structure, the network deconfounder suffers the least when the influence of hidden confounders grows (from $\kappa_2=0.5$ to $\kappa_2=2$) in terms of the increase in the errors $\sqrt{\epsilon_{PEHE}}$ and $\epsilon_{ATE}$.
\end{itemize}

\begin{table*}[tbh!]
\centering
\small
\caption{Experimental Results comparing effectiveness of the proposed network deconfounder with the baseline methods.}\label{tab:res1}
\begin{tabular}{|c|c|c|c|c|c|c|}
\hline
\multicolumn{7}{|c|}{BlogCatalog}\\\hline
$\kappa_2$ & \multicolumn{2}{|c|}{0.5}& \multicolumn{2}{|c|}{1}& \multicolumn{2}{|c|}{2}\\ \hline
 & $\sqrt{\epsilon_{PEHE}}$ &$\epsilon_{ATE}$ & $\sqrt{\epsilon_{PEHE}}$ &$\epsilon_{ATE}$ & $\sqrt{\epsilon_{PEHE}}$ &$\epsilon_{ATE}$\\
\hline
NetDeconf (ours) & \bf{4.532} & \bf{0.979} & \bf{4.597} & \bf{0.984} & \bf{9.532} &  \bf{2.130}
\\\hline
CFR-Wass & 10.904 & 4.257 & 11.644 & 5.107 & 34.848 & 13.053 \\ \hline
CFR-MMD & 11.536 & 4.127 & 12.332 & 5.345 & 34.654 & 13.785 \\ \hline
TARNet & 11.570 & 4.228 & 13.561 & 8.170 & 34.420 & 13.122  \\ \hline
CEVAE & 7.481 & 1.279 & 10.387 & 1.998 & 24.215 & 5.566 \\ \hline
Causal Forest &  7.456 & 1.261 & 7.805 & 1.763 & 19.271 & 4.050 \\ \hline
BART & 4.808 & 2.680 & 5.770 & 2.278 & 11.608 & 6.418 \\ \hline
\end{tabular}

\begin{tabular}{|c|c|c|c|c|c|c|}
\hline
\multicolumn{7}{|c|}{Flickr}\\\hline
% $\kappa_2$ & \multicolumn{2}{|c|}{0.5}& \multicolumn{2}{|c|}{1}& \multicolumn{2}{|c|}{2}\\ \hline
 & $\sqrt{\epsilon_{PEHE}}$ &$\epsilon_{ATE}$ & $\sqrt{\epsilon_{PEHE}}$ &$\epsilon_{ATE}$ & $\sqrt{\epsilon_{PEHE}}$ &$\epsilon_{ATE}$\\
\hline
NetDeconf (ours) & \bf{4.286} & \bf{0.805} & \bf{5.789} & \bf{1.359} & \bf{9.817} &  \bf{2.700}
\\\hline
CFR-Wass & 13.846 & 3.507 & 27.514 & 5.192 & 53.454 & 13.269 \\ \hline
CFR-MMD & 13.539 & 3.350 & 27.679 & 5.416 & 53.863 & 12.115 \\ \hline
TARNet & 14.329 & 3.389 & 28.466 & 5.978 & 55.066 & 13.105  \\ \hline
CEVAE & 12.099 & 1.732 & 22.496 & 4.415 & 42.985 & 5.393  \\ \hline
Causal Forest & 8.104 & 1.359 & 14.636 & 3.545 & 26.702 & 4.324 \\ \hline
BART & 4.907 & 2.323 & 9.517 & 6.548 & 13.155 & 9.643 \\ \hline
\end{tabular}
\end{table*}

\begin{figure}[tb]
\centering
\begin{minipage}{0.35\textwidth}
\centering
\subfigure[$\sqrt{\epsilon_{PEHE}}$ of BlogCatalog ($\kappa_2=1$)]
{\label{fig:BCPEHE}\includegraphics[width=\textwidth]{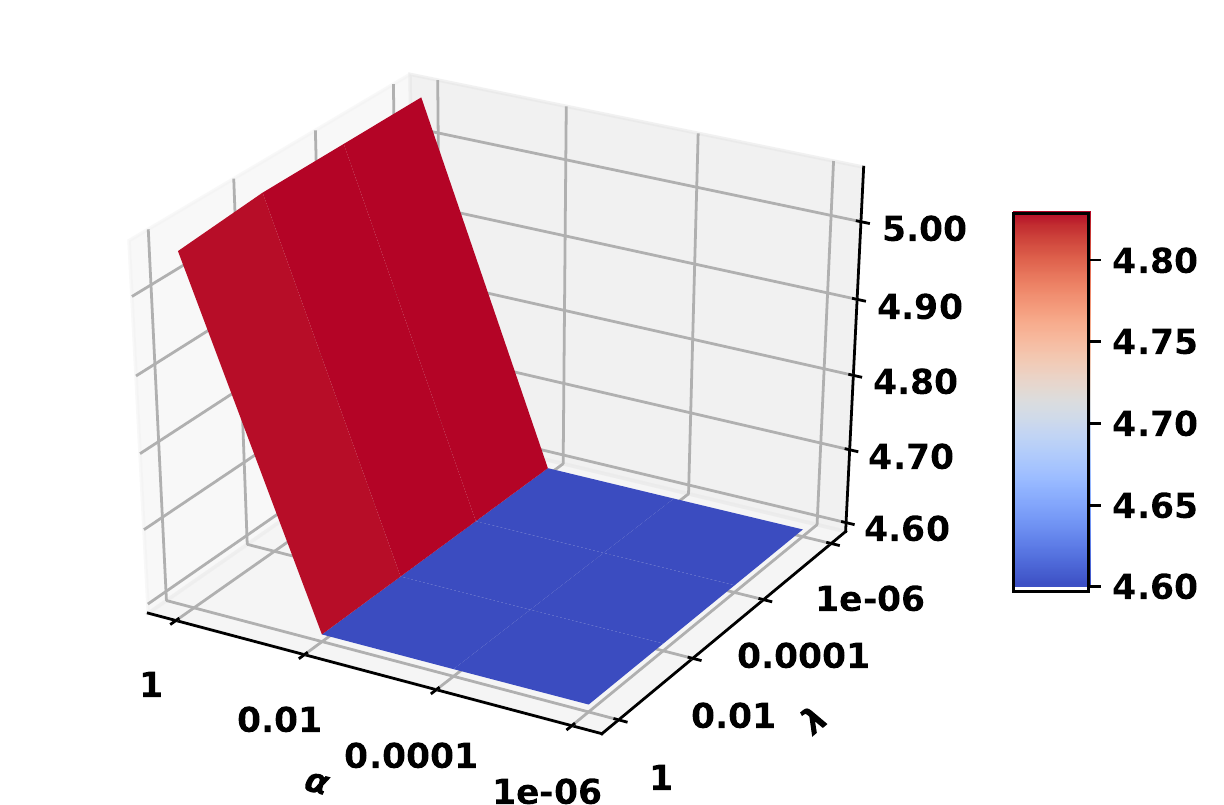}}
\end{minipage}
\begin{minipage}{0.35\textwidth}
\centering
\subfigure[$\epsilon_{ATE}$ of BlogCatalog ($\kappa_2=1$)]
{\label{fig:BCATE}\includegraphics[width=\textwidth]{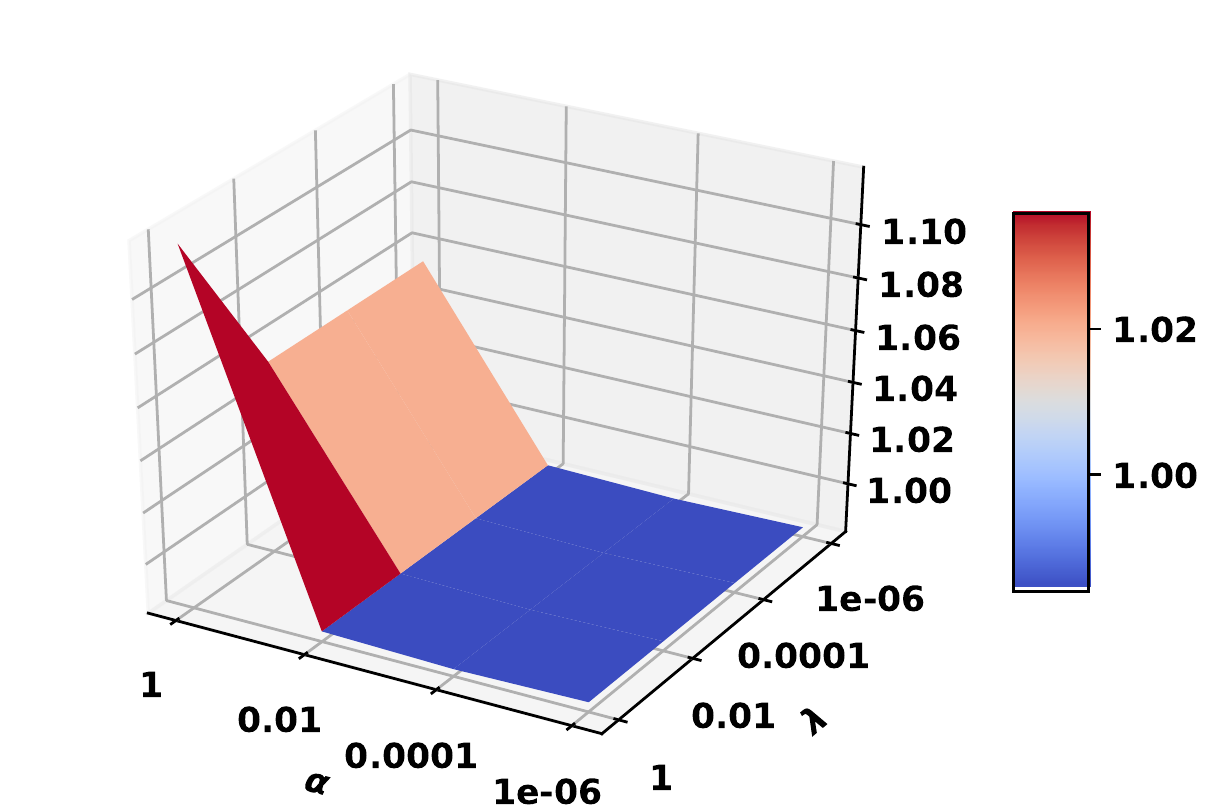}}
\end{minipage}
\centering
\caption{Parameter study results: impact of $\alpha$ and $\lambda$ on the performance of the proposed network deconfounder.}
% \vspace{-0.2in}
\label{fig:param_study}
\end{figure}

\noindent\textbf{Parameter Study.} Then we investigate how the values of the two hyperparameters $\alpha$ and $\lambda$ affect the performance of the network deconfounder.
Regarding to the settings for the parameter study, we fix the learning rate to be $10^{-2}$, the number of epochs to be 200, the number of GCN layers and the number of output layers to be $2$, the number of hidden units and the dimensionality of the representations to be $100$.
Then we vary $\alpha$ and $\lambda$ in the range of $\{1, 10^{-2}, 10^{-4}, 10^{-6}\}$.
The results are shown in Fig.~\ref{fig:param_study}.
Due to the space limit, we only report the results on the BlogCatalog dataset with $\kappa_2=1$ in terms of both error metrics $\sqrt{\epsilon_{PEHE}}$ and $\epsilon_{ATE}$.
Based on the observations that the $\sqrt{\epsilon_{PEHE}}$ and $\epsilon_{ATE}$ do not change significantly when $10^{-6}\le \alpha\le 10^{-2}$ and $10^{-6} \le \lambda \le 1$, we can conclude that the performance of the network deconfounder is not sensitive to both hyperparameters $\alpha$ and $\lambda$. However, when $\alpha$ is too large ($\alpha > 10^{-2}$), the performance of the network deconfounder degrades. This is because when $\alpha$ is too large, the objective function would emphasize the importance of balancing the confounders' representations of the two treatment groups too much and sacrifice the precision on the inferred ITEs.

\section{Related Work}
In this section, we present two directions of related work: learning individual level causal effects from observational data (without network information) and graph convolutional neural networks.

\noindent\textbf{Learning Individual Treatment Effects.}
Recently, the fundamental problem of learning ITEs from observational data has been attracting considerable attention in a myriad of applications.
Hill~\cite{hill2011bayesian} proposed to apply the regression  model BART~\cite{chipman2010bart} to estimate treatment effects. 
Then BART is widely adopted for causal effect estimation because it only requires little effort in hyperparameter tuning, provides posterior distributions of inferred outcomes for uncertainty quantification and can deal with both discrete and continuous treatment variables~\cite{hahn2017bayesian}.
Wager and Athey~\cite{wager2018estimation} proposed the Causal Forest, which extended the original random forest~\cite{breiman2001random} for estimating heterogeneous treatment effects, which is the expected ITE over the individuals with the same features.
This method partitions the original feature space into subspaces. In each subspace, the instances are very similar to each other such that their treatment assignments can be considered as randomized.
In~\cite{johansson2016learning,shalit2017estimating}, two methods are proposed to apply representation learning techniques to controlling confounding bias for causal inference. 
Theoretically, Shalit et al.~\cite{shalit2017estimating} showed that balancing the distributions of the treated and controlled instances in the representation space of confounders can be helpful in the task of estimating ITEs.
However, the existing methods mentioned above rely on the strong ignorability assumption which essentially ignores the influence of hidden confounders, which is often untenbale and can be unrealistic in real-world observational studies.
Louizos et al.~\cite{louizos2017causal} proposed to consider observed features as proxy variables of confounders and developed a deep latent-variable model to learn representations of confounders through variational inference.
However, none of the previous work utilized the network structure to capture patterns of hidden confounders.

\noindent\textbf{Graph Convolutional Networks.}
Previous work on Graph Convolutional Networks (GCN) mainly focused on the development of spatially localized\footnote{Here, spatial locality refers to the constraint that information of a node only propagates to its neighbors in a certain number of hops.} and computationally efficient convolutional filters for various types of network data including citation networks and social networks.
Bruna et al.~\cite{bruna2013spectral} proposed to use the first-order graph Laplacian matrix as the basic of filters in the spectrum domain.
However, this filter has a large number of trainable parameters and its the spatial locality is not guaranteed.
In~\cite{defferrard2016convolutional}, Defferrard et al. proposed a more efficient and properly localized filter for the graph convolution operator. This filter is parameterized as $l$-th order polynomials of the graph Laplacian matrix to ensure the locality, where $l$ is a positive integer and is often greater than $1$. 
Then the polynomials are approximated by their Chebyshev expansion to reduce the computational cost.
Then, Kipf and Welling~\cite{kipf2016semi} proposed the renormalization trick to further improve the computational efficiency of GCN.
Recently, variants of GCN has also been proposed to a myriad of applications using network data such as recommendation~\cite{wang2019neural}, content recommendation in social networks~\cite{ying2018graph}, anomaly detection in attributed networks~\cite{ding2019deep}, entity classification and link prediction in knowledge graphs~\cite{schlichtkrull2018modeling} and link sign prediction in signed networks~\cite{derr2018signed}.
Different from the existing work, this paper is the first work exploiting GCN for causal inference with observational data.
\section{Conclusion}
New challenges are presented by the prevalence of networked observational data for learning individual treatment effects.
In this work, we study a novel problem, learning individual treatment effects from networked observational data.
As the underlying network structure could capture useful information of hidden confounders, we propose the network deconfounder framework, which leverages the network structural patterns along with original features for learning better representations of confounders.
Empirically, we perform extensive experiments across multiple real-world datasets. Results show that the network deconfounder learns better representation of confounders than the state-of-the-art methods.
%
%This is because it utilizes the graph convolutional neural network layers to recognize patterns of hidden confounders from the network structure.
%

Here, we also introduce two most interesting directions of future work. 
First, we are interested in leveraging other types of structure between instances for learning ITEs from observational data. For example, temporal dependencies can also be utilized to capture patterns of hidden confounders.
Second, in this work, we focus on static network structure. But the real-world networks can evolve over time~\cite{marin2017temporal,sarkar2019using}. Hence, we would like to investigate how to exploit dynamics in evolving networks for learning ITEs.

% \clearpage

\section*{Acknowledgement}
This material is based upon work supported by ARO/ARL and the National Science Foundation (NSF) Grant \#1909555.

% \pagebreak

\bibliographystyle{ACM-Reference-Format}
\bibliography{acm_main}
\end{document}